# The FourStar Infrared Camera

S. E. Persson[1], Robert Barkhouser[2], Christoph Birk[1], Randy Hammond[2], Albert Harding[2], E. R. Koch[3], J. L. Marshall[1], Patrick J. McCarthy[1], David Murphy[1], Joe Orndorff[2], Gregg Scharfstein[2], Stephen A. Shectman[1], Stephen Smee[2], Alan Uomoto[1]

[1]Observatories of the Carnegie Institution of Washington
813 Santa Barbara Street, Pasadena, CA 91101

[2]Instrument Development Group, Johns Hopkins University
3400 N. Charles St., Baltimore, MD 21218

[3]Dedicated Micro Systems
3095 N. E. Nathan Dr., Bend, OR 97701

## ABSTRACT

The FourStar infrared camera is a 1.0-2.5 µm (JHK$_s$) near infrared camera for the Magellan Baade 6.5m telescope at Las Campanas Observatory (Chile). It is being built by Carnegie Observatories and the Instrument Development Group and is scheduled for completion in 2009. The instrument uses four Teledyne HAWAII-2RG arrays that produce a 10.9' x 10.9' field of view. The outstanding seeing at the Las Campanas site coupled with FourStar's high sensitivity and large field of view will enable many new survey and targeted science programs.

## 1. INTRODUCTION

The FourStar infrared camera is a new, large field of view JHK$_s$ camera for the Magellan Baade telescope. The outstanding image quality of Magellan and the high performance and wide field of FourStar will enable a range of survey and targeted programs. The instrument is currently in the build phase. The mechanical hardware has been fabricated, the control electronics are complete, the detectors are being characterized, and the coding of the data acquisition system is being finalized. Figure 1 shows a cutaway view of the instrument.

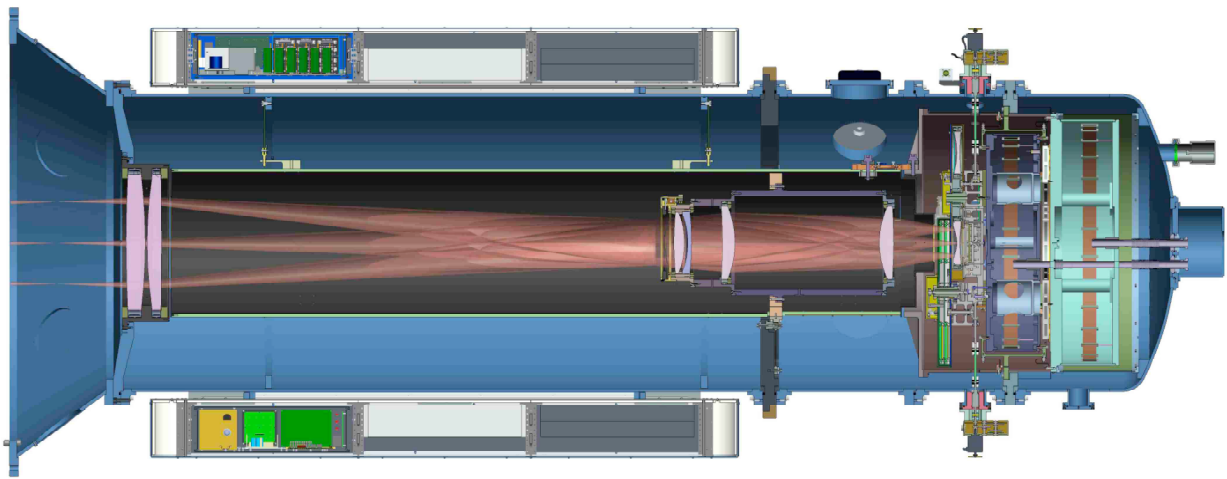

**Figure 1.** Mechanical drawing of the FourStar instrument.

## 1.1. Instrument Overview

FourStar provides a 10.9 arcminute square field of view using a 2x2 array of Teledyne HAWAII-2RG focal plane arrays that provide a plate scale of 0.159 arcsec per pixel. This scale slightly under-samples images under the best seeing conditions at Magellan, which have been measured with the PANIC camera on the Magellan Baade telescope (Martini et al. 2004), to be at best 0.21 arcsec FWHM at $K_s$ in ten minutes and a median of 0.41 arcsec FWHM.

| *Property* | *Value* | *Notes* |
|---|---|---|
| Focal Plane Format | 4096 x 4096 | Four HAWAII-2RG arrays |
| Pixel Scale | 0.159″ | 18μm pixels |
| Field of View | 10.9′ x 10.9′ | |
| Sensitivity | $K_s$ = 20.6 (Vega) | 5σ point source in 1hr |

**Table 1.** Key FourStar parameters.

FourStar will be outfitted with a standard suite of $JHK_s$ filters. The instrument has two 6-position filter wheels in series, and may accommodate up to 10 filters. The remaining filters will be designed for particular science projects, and will have narrow ($\lambda/\Delta\lambda\sim100$) or medium ($\lambda/\Delta\lambda\sim10$) bandwidths.

The instrument will reside at one of the two *f*/11 Nasmyth ports of the Baade 6.5m telescope. It is about 3 m long and weighs 1200 kg. Most of its weight will be supported by the instrument handling cart which sits on the Nasmyth platform; attachment to the telescope via the guider is used only to locate the instrument in space.

## 1.2. Science with FourStar

FourStar will be primarily a survey instrument. Its combination of sensitivity, image quality and field of view are ideally suited to addressing key problems in the study of distant galaxies, star formation, and stellar astrophysics.

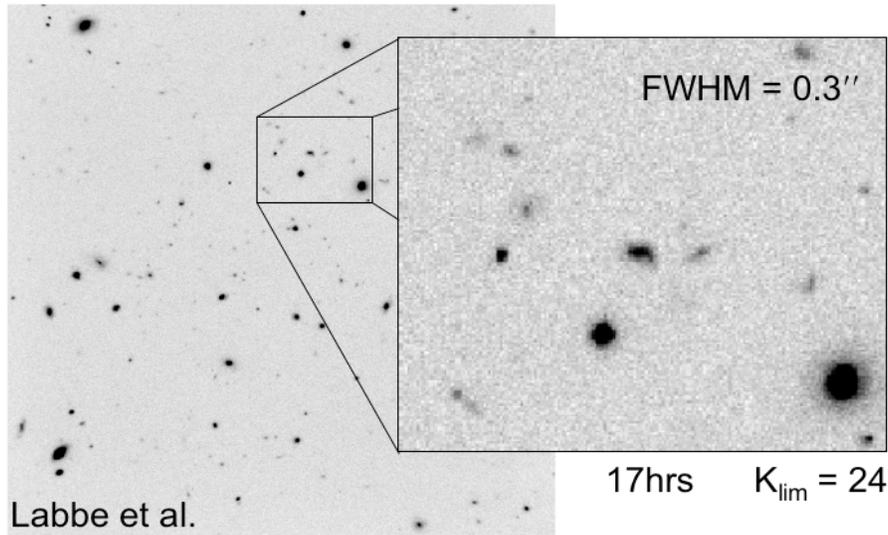

**Figure 2.** An example of the image quality and sensitivity offered by the Magellan Telescopes. This 17 hour Ks-band image of the Hubble Ultra-Deep Field with PANIC (Martini et al. 2004) reaches a limiting depth of $K_s$ = 24 mag (Vega). The resolution of the images rivals that of NICMOS on HST. The 25 times larger field of view of FourStar will make deep surveys over moderate areas practical and will enable studies of massive red galaxies in the 1 < z < 3 range with high precision and statistical accuracy. The image is from Labbé (2007).

The highest priority scientific programs for FourStar center on studies of the distant universe. Near-IR imaging surveys allow the selection of galaxies on the basis of their stellar mass and tracing the evolution of rest-frame visible light to early epochs. Carnegie astronomers plan to carry out surveys designed to study the assembly of stellar mass in galaxies in the $1 < z < 3$ epoch. These surveys will be of intermediate area and depth and will build on extant data at other wavelengths. The near-IR photometry aids determinations of stellar mass and star formation histories in massive galaxies, particularly passively evolving systems, as illustrated in Figure 2. At higher redshifts, ultra-deep surveys allow us to trace the evolution of the most massive galaxies in the $3 < z < 4$ era when the first massive stellar systems appeared. FourStar will complement the capabilities of the IMACS and LDSS3 spectrographs on the Magellan telescopes and will allow powerful surveys over a wide range of redshifts. Through the use of medium-band filters in the J and H bands we plan to carry out deep photometric redshift surveys that will have higher precision than those afforded by only broad band observations.

## 2. OPTICAL DESIGN

The FourStar optical design is similar to that of the PANIC infrared camera (Martini et al. 2004); see Figure 3. The instrument has seven elements in three groups. The two-element vacuum vessel entrance window (L1 & L2) is also a Fabry lens that images the telescope entrance pupil onto an adjustable-diameter cold stop inside the instrument, just in front of lens L3. The choice of a doublet for L1/L2 reduces the chance of forming condensation on the dewar window (L1) by keeping the center of the vacuum window warmer. Four lenses (L3-L6) and a field flattener (L7) bring the beam to focus at $f/3.6$. There is a different field flattener lens for each of the J, H, and $K_s$ bands. One aspheric surface (on L3) is needed to obtain the large field area. The "camera module" optics, L3-L6, are cooled to 200K, a temperature low enough to ensure negligible background contribution from the instrument. The mechanisms, filters, field flattener lenses, and detectors near the focal plane are cooled to 77K. Two filter wheels can accommodate up to 10 filters. The final plate scale delivered by the $f/3.6$ beam is 0.159 arcsec/pixel accommodating a total field of 10.9' X 10.9'.

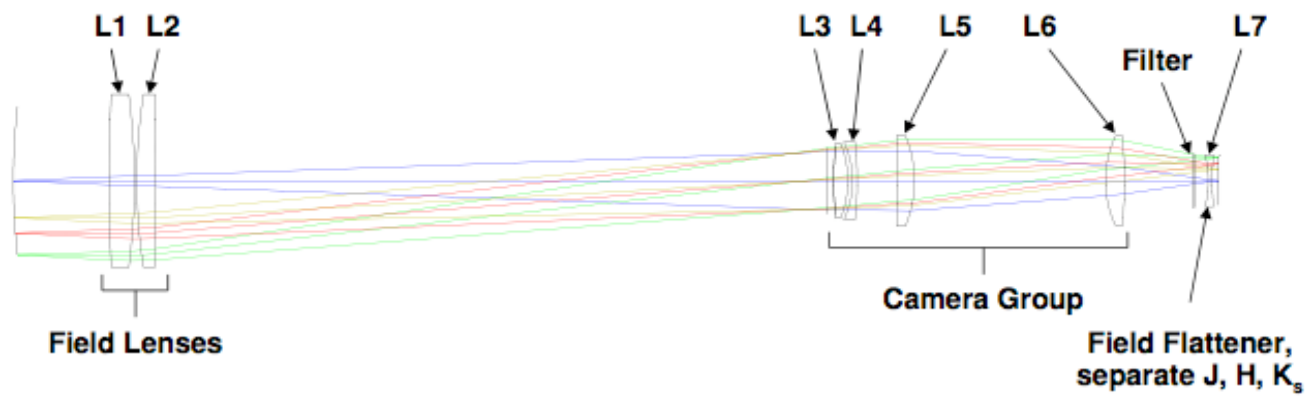

**Figure 3.** FourStar optical design.

## 3. MECHANICAL DESIGN

### 3.1. Vacuum vessel

The cylindrical aluminum vacuum vessel forms the primary instrument structure (see Figure 1). It provides a vacuum environment sufficient to eliminate convection and gas conduction heat losses between the warm ambient environment and the cold optics. The aluminum vessel is 900 mm in diameter and 3 m in length. It is comprised of four separable sections to facilitate assembly, optical alignment, and testing. The entrance end of the vessel is capped by a dished cover; the central hole serves as the opto-mechanical interface for window assembly. A thick rigid ring between the first two sections supports the camera module (L3-L6), and a second ring between the last two sections supports the focal plane dewar. Viton O-rings seal the sections and precisely machined stepped interfaces guarantee alignment tolerances are maintained if the sections are separated and reassembled. Numerous penetrations exist for electrical and mechanical feedthroughs, as well as access ports to ease assembly. An ion pump and activated charcoal getter pump the volume under steady state operation.

### 3.2. Optical mounts

FourStar consists of three opto-mechanical sub-assemblies. The window assembly contains L1 & L2, the camera module contains L3 - L6, and the three field flatteners (L7) – one for J, H, and $K_s$ – and filters are mounted in mechanized wheels just in front of the focal plane. The grouping of the various elements arose naturally from their physical location within the system and their sensitivity to misalignment.

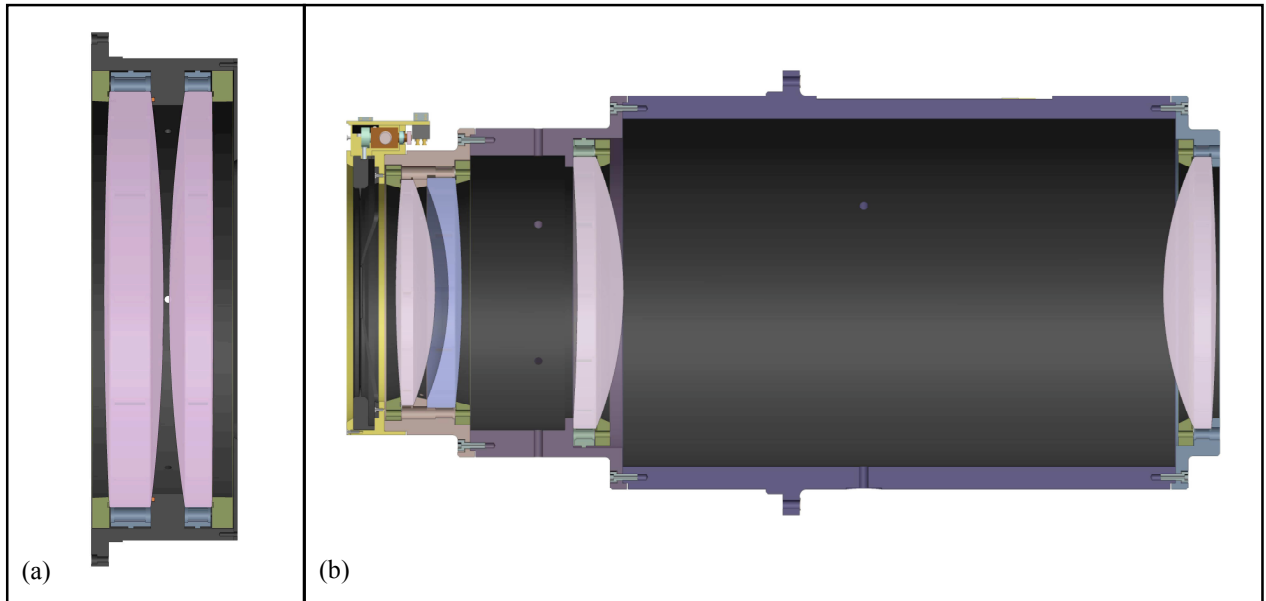

**Figure 4.** Cross sections of: (a) the window assembly, and (b) the camera module.

Figure 4 shows a cross section of the window and the camera module. Lens centration is achieved using 7075-T6 aluminum cells that contain precisely machined roll-pin flexures (Smee 2008) that center the lens and provide radial compliance to accommodate the differential thermal expansion between the glass and the lens cell. Axial placement is defined by a stepped interface built into the cell or barrel. A retention ring is preloaded against the face of the lens using spring washers, which firmly seats the lens and accommodates coefficient of thermal expansion (CTE) mismatches. Lens mating surfaces are anodized with a Polytetrafluoroethylene (PTFE) coating applied as a post process to reduce the coefficient of friction at the metal/glass interface. The machined components are aluminum, which reduces fabrication cost and mass. The low mass is a significant benefit as it reduces image motion caused by gravity-induced movement of the camera module with respect to the incoming beam as the Nasmyth axis rotates.

The field flatteners are also mounted in roll-pin flexure cells. Figure 5 shows an exploded view of the L3/L4 cell, with a detail of the roll-pin flexure.

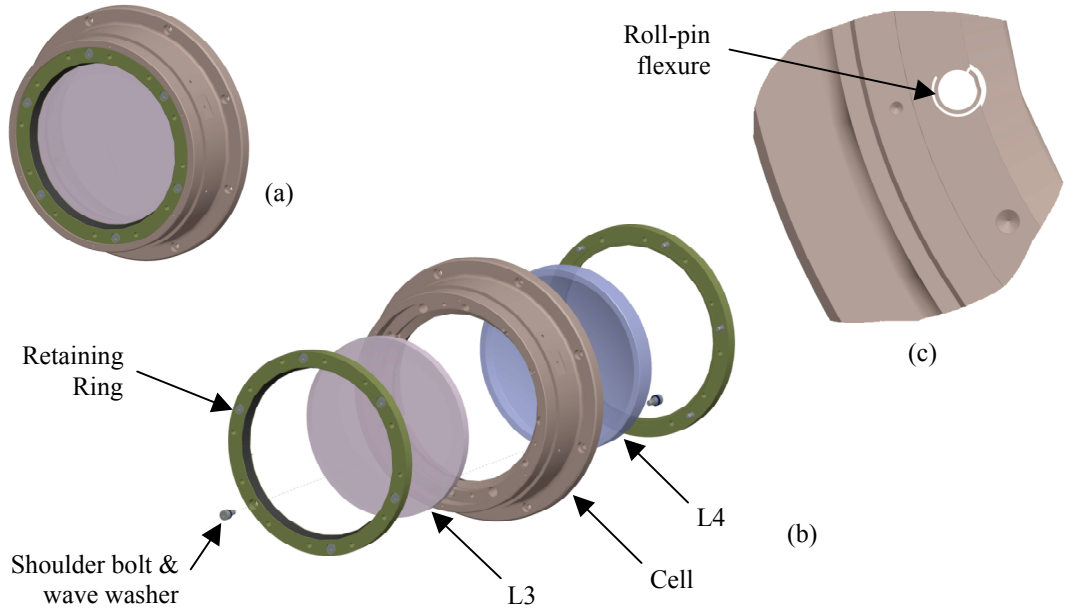

**Figure 5.** L3/L4 cell design. (a) Assembled cell, (b) exploded view detailing cell components, and (c) roll-pin flexure detail. Each lens is centered by six roll-pin flexures, which provide radial compliance during warmup and cooldown.

### 3.3. Alignment

An auto-collimating alignment telescope mounted to a precision interface on the front of the vacuum vessel defines the optical axis of the instrument; see Figure 6a. This interface is machined coaxial and parallel to the window interface. In this way, the window assembly, which has a loose centering/tilt tolerance, can be mounted to the vessel without adjustment. A plate containing a small window that allows use of the collimating telescope is substituted for the L1/L2 cell when cold measurements are taken.

The camera module support ring at the opposite end of the first section is centered and squared to the alignment telescope axis; a precision target on the support ring serves as a fiducial for the alignment process. In a similar fashion a target in the field flattener wheel is used to align the field flattener with the optical axis. Precision metrology and shims are used to adjust the vertex-to-vertex spacings between the window and camera module, as well as the spacing between the camera module and field flattener; the tolerances on these vertex spacings are not tight (~ ± 0.5 mm).

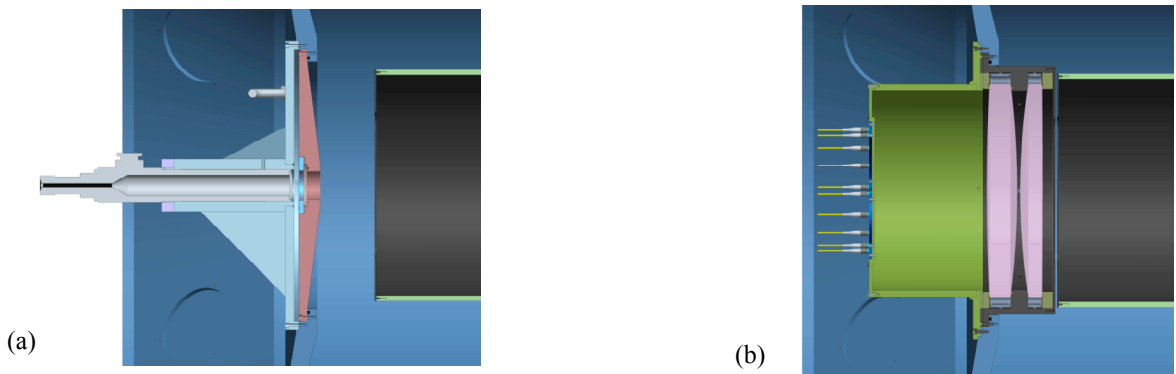

**Figure 6.** Optical alignment fixtures. (a) The alignment telescope mounts to the front of the vessel and establishes the optical axis, and (b) the fiber array produces a simulated star pattern used for alignment of the detector array.

To align the detector module, the optical center of the cold work surface is located with the alignment telescope and kinematic mounting feet are machined so as to achieve the correct XY position. Tip, tilt, and piston are adjusted using three fine-threaded adjustment screws that engage the feet.

The detector module has an adjustable focus that is used to obtain a set of cold through-focus measurements of an array of fibers positioned at the nominal telescope focus just in front of the window. These measurements are used to determine how much tip, tilt, and piston adjustment is required; see Figure 6b.

### 3.4. Thermal Regimes

Two cold regimes are used in FourStar. Both are established using liquid nitrogen contained in two dewars positioned in series at the back end of the instrument.

The rearward dewar (i.e., the shield dewar) provides the primary cooling power for the instrument. It is supported by a thin-walled G10 tube connected to the rear of the vacuum vessel. It cools both the 77 K radiation shield, which encloses the focal plane, and the 200 K shield, which surrounds the camera module and extends up to the entrance window. The temperature difference between the two regimes is established by a set of thermal resistors that bridge the two radiation shields. The thermal load on the dewar is roughly 60 W, its capacity is 30 liters, and its hold time is about one day. An autofill system will supply $LN_2$ daily from a large external supply dewar.

The forward dewar (i.e., the detector dewar) cools the detector array and focal plane mechanisms. It is supported mechanically by G10 flexures and is surrounded by the 77 K radiation shield; hence the heat load is limited to conduction through the flexures and electrical cables and a small viewfactor to the camera module. This ensures a stable operating temperature for the array and mechanisms. The heat load on the detector dewar is a few watts, its capacity is 15 liters, and it should have a hold time of greater than one week. Unlike the shield dewar, it will be filled manually.

### 3.5. Focal Plane Mechanisms

The internal mechanisms of FourStar hold and locate the detectors (with cold focus ability), the field flattener and filter wheels, and resize the L3 cold stop for operation in the $K_s$ band. Figure 7a shows an exploded rendering of the cold work surface mechanisms. The three field flattener lenses are mounted in a wheel forward of the focal plane. A fourth position is used solely for optical alignment and contains the laser alignment test optic. There are two filter wheels, each with five filter spaces and one open position. Although FourStar was not particularly designed for narrow-band applications, narrow-band (~ 1%) or intermediate-band (~ 10%; e.g., $CH_4$) filters may be used.

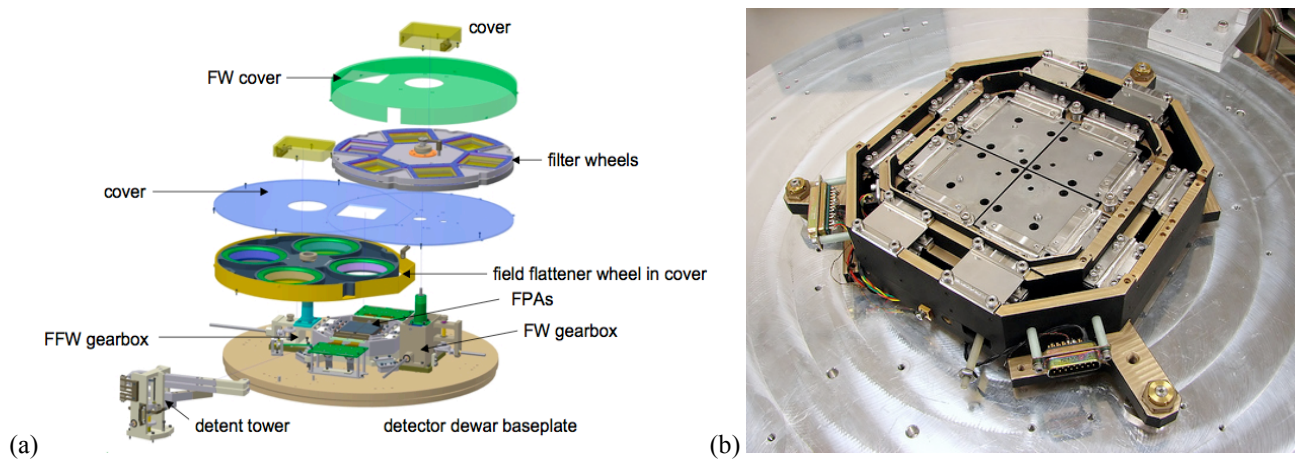

**Figure 7.** (a) Exploded rendering of the focal plane mechanisms. (b) Photograph of the FourStar detector mount.

Figure 7b shows a photograph of the detector module. The detector module design allows alignment of the detector focal plane orthogonal to the optical axis. The procedure for achieving detector co-planarity and optical axis orthogonality depends on detector module adjustments, used in concert with a specially designed optical alignment fixture/lens that images the laser "stars" onto the detectors (see Figure 6b). Each detector is mounted on its own molybdenum block, the blocks are mounted on the module, and the module itself is kinematically mounted on the detector dewar cold work surface.

### 3.6. Instrument Handling Carts

Two handling carts have been built for FourStar: the instrument cart and the back-end cart.

The instrument cart, shown in Figure 8, supports FourStar both on and off the telescope. On the telescope, the instrument cart supports the cantilevered load of the instrument through a spring-loaded yoke assembly that contacts the instrument load ring. The spring force is adjusted to tune the reaction force to minimize instrument flexure. The spring load is transferred to the load ring by a pair of rollers that allow free rotation of the Nasmyth axis. Translational degrees of freedom have been built into the yoke assembly to handle instrument runout.

Four leveling jacks provide rigid support both on the telescope and in the lab. The cart has a two-tiered structural design with ball transfers between the upper and lower cart halves. The ball transfers ease the mating of the instrument to the telescope by allowing low-friction fine alignment adjustments without the necessity of moving the instrument on its casters. In the laboratory, the cart serves as an integration and testing platform and provides a convenient mode of transport.

The back-end cart is used for the integration and testing of the dewars and focal plane mechanisms housed in the rear section of the instrument. This cart is also used to assemble the back end of the instrument to the front end. In serving both functions the cart must provide two nominal configurations: raised and lowered. In the lowered configuration the vessel sections are supported axis-vertical; the dewar work surface is horizontal and at a comfortable working height. In the raised configuration, the vessel sections are rotated to axis-horizontal and raised to the same elevation as the front half of the instrument.

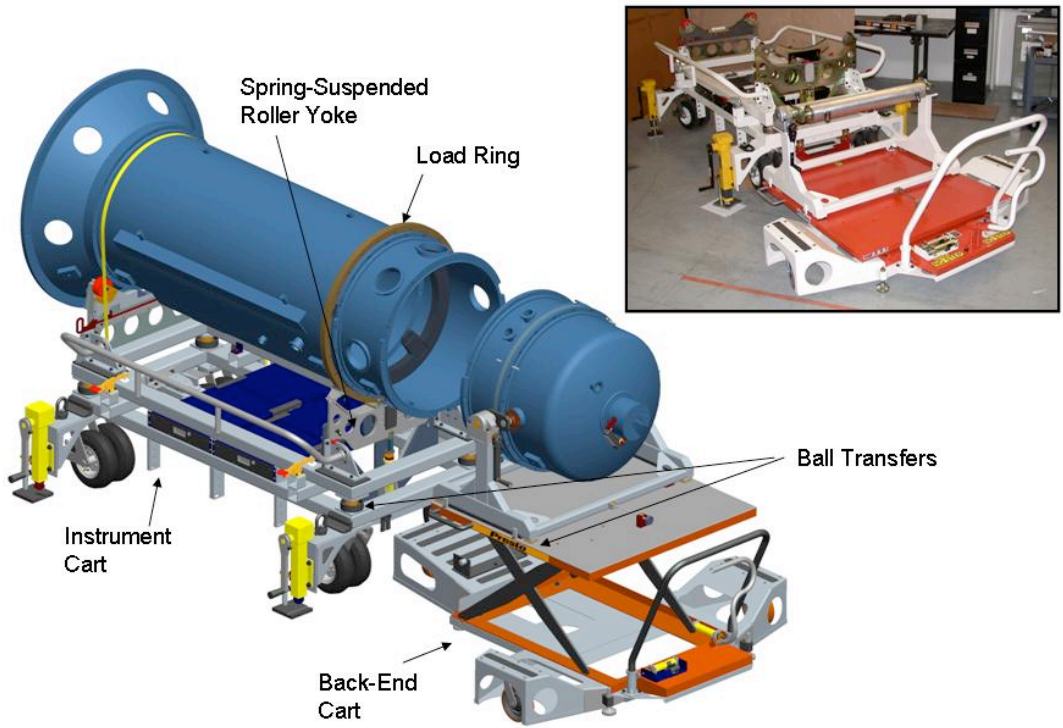

**Figure 8.** FourStar handling carts.

# 4. DETECTORS AND CONTROL ELECTRONICS

## 4.1. HAWAII-2RG Detectors

The FourStar detectors are four substrate-removed HAWAII-2RG 2048x2048 arrays manufactured by Teledyne Scientific and Imaging (formerly Rockwell Scientific) mounted in a 2x2 format. The arrays will be read out by the new application-specific integrated circuit (the "SIDECAR" ASIC) devices designed by Teledyne; each detector has its own associated ASIC. These arrays are sensitive into the optical wavelengths (blueward to ~5500 Å); hence the FourStar filters are designed to allow only infrared light to pass.

The Teledyne detector system consists of only three parts: the detector itself, the ASIC, and a "Jade2" card to couple signals from the detector to a computer via a USB 2.0 interface. Commands from the control computers and data from the detector/ASIC use this link. Teledyne provides a driver interface allowing communication with the ASIC. Following Teledyne's lead, PCs running Windows XP are used to control the ASICs and accept imaging data.

The mounting of two ASICs to serve abutted detectors presents a problem because the center-to-center spacings of the chips (ASIC vs. HAWAII-2RG) are not commensurate. The ASICs require more room, and thus a specially-designed board to hold two ASICs was developed. Teledyne has designed and delivered the boards, as well as the cables needed to connect them to the Jade2 cards. Figure 9 shows a photo of one channel of the detector hardware.

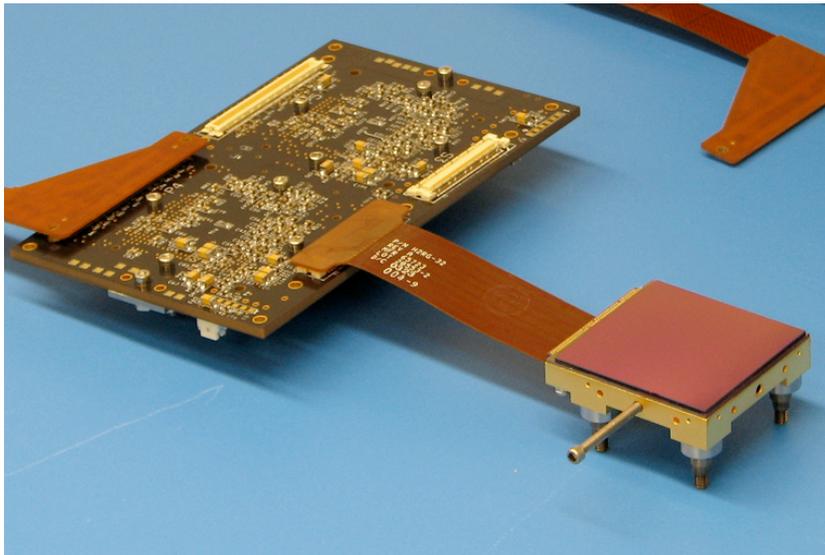

**Figure 9.** Photograph of the FourStar HAWAII-2RG engineering-grade array connected to one channel of a dual-channel ASIC board.

All four science-grade arrays have been delivered to Carnegie. Teledyne measurements indicate excellent quantum efficiency (QE), dark current, and pixel operability. The QE at 2 μm is approaching 100% and the J-band QE is 80%. Some uncertainty remains in the readnoise data. Teledyne measures ~30 electrons rms, which is out of spec by ~50%. The validity of this measurement is suspect due to test setup uncertainties. Further measurements will be required to determine the exact readnoise.

## 4.2. Data Acquisition System

The data acquisition system consists of four Windows PCs that read out the four arrays, and seven Macintosh computers that process the data and control the instrument. Five of these Macintoshes are devoted to real-time data processing, so that the observer will leave the mountain with fully reduced data. Figure 10 gives a diagram of the system.

The FourStar control- and data-acquisition software consists of two programs. A graphical user interface (UserGUI) and a data server (DataServer).

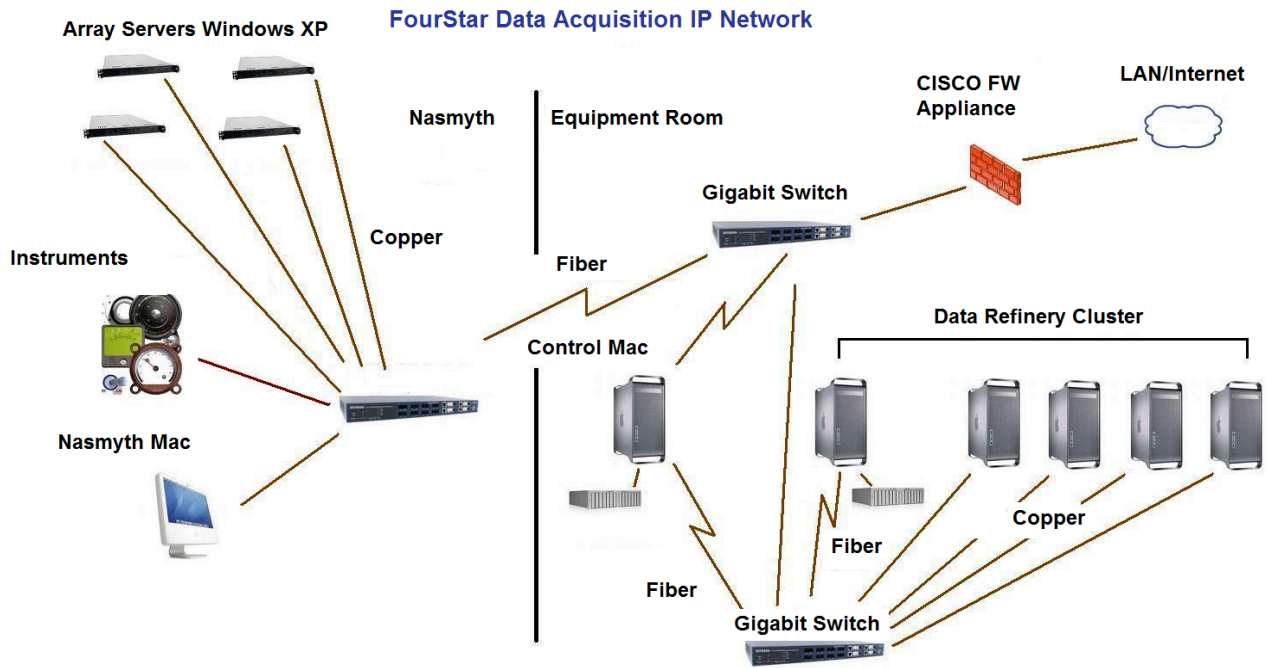

**Figure 10.** Diagram of the FourStar data acquisition IP network.

The DataServer program runs under Windows XP on four computers, each serving one focal plane array (FPA). Each FPA is controlled by an ASIC that communicates via USB with the server software. The ASIC programming allows for various readout modes, e.g. correlated double sampling (CDS). The DataServer receives the exposure parameters from the user GUI and reads the science data from the FPA. It preprocesses the raw data frames (e.g., it subtracts reference pixels) and delivers the resulting images to the UserGUI. The communication with the user GUI is done via a TCP/IP connection.

The UserGUI program runs on a Mac computer. It controls the motion stages (filter-wheel, focus, etc) and supervises the instrument temperature controllers and monitors. It allows the user to set the exposure parameters and sends them to each of the four data servers. It writes the science data as FITS files to disk and shows the images on a quick-look display in real time. The UserGUI also communicates with the telescope control system thereby enabling automatic dither sequences (macros).

### 4.3. Instrument Electronics

Two electronics racks mounted on the instrument contain the chassis required to monitor and control the various functions of the instrument (see Figure 1). These include the four data acquisition computers, motor controllers for the instrument mechanisms, the liquid nitrogen level monitoring/control system, and the temperature monitoring/control systems. The racks are enclosed and insulated; heat is carried away via a glycol/water heat exchange system.

## 5. DATA REDUCTION

Under normal weather and $K_s$ background conditions, the amount of raw data accumulated during a night ($K_s$ observations only) could exceed 100 Gigabytes. These are 16-bit CDS frames, taken at the rate of 4 $min^{-1}$. For a several-night run, writing this much raw data on mass media will require an inordinate amount of time. It is essential, therefore, to develop a pipeline to reduce the data during the night and into the next day, such that the observer can leave the mountain with data reduced to final form. We have created prototype code and used simulated data to assess these requirements quantitatively. In addition to the main reduction task, we will operate a subset of the full reduction steps in quasi-real time, for the observer to assess the data quality and conditions (seeing, transparency, background). The results of these simulations indicate that achievable code optimization must be done, but that this is a reasonable goal.

## 6. CURRENT STATUS AND INSTRUMENT DEPLOYMENT

FourStar has been completely designed and we are currently in the process of assembling the mechanical hardware in the lab. We have received shipment of the vacuum vessel sections, the two dewars, the detectors and readout electronics, all of the optics, instrument handling carts, and all of the internal mechanisms. The instrument control software and detector control software are being finalized and the data reduction pipeline is underway. The remainder of 2008 will be spent assembling and testing the instrument in the Class 10,000 clean room at OCIW. We expect to ship FourStar to Las Campanas in 2009.

## ACKNOWLEDGEMENTS

Support was provided by NSF/ATI Grant #0138278 in the form of providing sufficient funding to acquire the four HAWAII-2RG detectors.